\documentclass[nonacm]{acmart}
\AtBeginDocument{%
  }
\setcopyright{none}

\author{Muhammad Raees}
\email{mr2714@rit.edu}
\orcid{0000-0002-1581-0378}
\affiliation{%
  \institution{Rochester Institute of Technology}
  \city{Rochester}
  \state{New York}
  \country{USA}
}

\author{Konstantinos Papangelis}
\email{kxpigm@rit.edu}
\orcid{0000-0001-5094-9837}
\affiliation{%
    \institution{Rochester Institute of Technology}
    \city{Rochester}
    \state{NY}
    \country{USA}}

\usepackage{multirow}

\newif\ifshowchanges 
\showchangestrue 
\newcommand{\changed}[1]{%
    \ifshowchanges 
        #1%
    \else 
        #1%
    \fi
}

\usepackage[most]{tcolorbox}
\usepackage[table,xcdraw]{xcolor}
\definecolor{lightbluebg}{HTML}{E6F0FF}
\definecolor{ribbonblue}{HTML}{4A90E2}

\newcommand{\custombox}[1]{
  \begin{tcolorbox}[
    colback=lightbluebg,        
    colframe=white,             
    boxrule=0pt,
    left=2mm,
    borderline west={2mm}{0pt}{ribbonblue},  
    sharp corners,
    before skip=10pt,
    after skip=10pt,
    enhanced
  ]
    #1  
  \end{tcolorbox}
}

\begin{document}

\title[From Trust to Appropriate Reliance]{From Trust to Appropriate Reliance: Measurement Constructs in Human-AI Decision-Making}

\renewcommand{\shortauthors}{Raees and Papangelis}

\begin{abstract}
While human-AI decision-making research has primarily used trust measurements to assess the practical usage of AI systems by their end-users, recent empirical evidence suggests that trust measurements do not inform users' \textit{appropriate reliance} on AI systems. 
While examining the human-AI decision-making literature, in this work, we review empirical studies that assess people's appropriate reliance on AI advice, differentiating measurements and constructs of appropriate reliance from trust and mere reliance.
Our analysis of literature shows that constructs for human-AI appropriate reliance are still fragmented in research. 
We present three views on appropriate reliance, namely Traditional, Appropriateness, and Dominance, as discussed in research. 
Using these views, we evaluate objective metrics reported in studies and argue for their consensus to facilitate the comparison across empirical research.
We also discuss how studies employ objective metrics and examine their validity in application contexts.
Our work contributes to the critical body of research on exploring objective metrics for assessing humans' appropriate reliance on AI advice.
\end{abstract}


\maketitle

\section{Introduction}
\label{sec-intro}
Artificial Intelligence (AI) systems assist humans in their decision-making in diverse domains, including healthcare, education, and business~\cite{dellermann2019hybrid, corbett2017algorithmic, longo2024explainable, green2019principles}.
However, concerns about AI's transparency, accountability, and improper usage have also sparked criticism for its \textit{misuse and disuse}~\cite{cabitza2023ai, lai2023towards, mckinney2020international}.
To this end, research has explored various means (e.g., explanations~\cite{ali2023explainable}, interaction~\cite{raees2024explainable}) to enhance transparency, accountability, and proper usage of AI systems; yet humans often over- or under-rely on AI, leading to inferior performance on decision-making tasks~\cite{green2019principles, bansal2021does}.
To reduce over- or under-reliance on AI systems, humans should only rely on AI when it provides correct decision recommendations and not rely on it when it provides incorrect decision recommendations~\cite{schemmer2023appropriate}. 
In research, this notion of reliance is often termed as \textit{``appropriate reliance''}~\cite{schemmer2023appropriate}.


Building reliance on AI-assisted decision support is rooted in building \textit{trust} on it~\cite{scharowski2023exploring, siau2018building, lankton2015technology}.
Commonly, users' trust in a system is measured as 1) subjective feedback (perceived trust), 2) behavioral assessment (demonstrated trust), and 3) a mix of both (actual/comprehensive trust)~\cite{mehrotra2024systematic, wischnewski2023measuring}.
For instance, users' trust can be measured as their perception of system performance and their inclination to rely on it~\cite{lee2004trust}.
\textit{``Appropriate trust''}, however, is linked with aligning users' perception with the actual system performance~\cite{mehrotra2024systematic, yang2020how}.
Users' reliance on AI, on the other hand, is demonstrated and stems from humans' \changed{tendency to adopt AI to achieve a task or make a decision~\cite{lai2023towards}.
Appropriate reliance is when humans only adopt AI when it provides correct advice or when they can objectively discriminate between the correct and incorrect AI advice~\cite{schemmer2023appropriate}.}
Existing work~\cite{mehrotra2024systematic, wischnewski2023measuring, lai2023towards, eckhardt2024survey} has explored the distinction between trust and reliance; however, confusion exists for evaluating human-AI reliance \changed{itself from appropriate reliance}.
In addition, no consensus exists for metrics for measuring appropriate reliance, which is used interchangeably with appropriate trust, \changed{which includes a degree of belief and personal feeling about the system alongside its actual performance}~\cite{mehrotra2024systematic}. 

In research, the metrics of measuring trust and reliance are studied~\cite{wischnewski2023measuring, eckhardt2024survey}; still, understanding \textit{appropriate reliance} and its measurements and constructs has been underexplored.
Such under-exploration provides a fragmented understanding of human-AI appropriate reliance research. 
Hence, in this work, we focus on nascent constructs and metrics used for assessing humans' appropriate reliance on AI advice, \changed{distinguishing them from metrics used for measuring mere AI reliance and trust.} 
We investigate the challenges of accurate assessment of users' objective performance for appropriate reliance on AI systems~\cite{lai2023towards, eckhardt2024survey, schemmer2023appropriate}. 
By doing so, \changed{we aim to provide a summary of constructs applied in the literature and discuss related research trends.}
We drive our investigation by exploring the following research questions;

\begin{enumerate}
    \item How does current research define appropriate reliance and related concepts?
    \item What and how are appropriate reliance constructs established, operationalized, and assessed in human-AI decision-making research?
\end{enumerate}

We attempt to answer these questions by conducting a review of recent empirical research on \changed{human-AI decision-making, studying appropriate reliance}. 
We first clarify the distinction between trust and reliance, and their conceptualization, providing a synthesis of related work. 
Then, we examine the literature that studies \changed{\textit{human-AI appropriate reliance}} in human-AI decision-making settings. 

Our findings show that research aims to assess human-AI appropriate reliance using several metrics, i.e., by measuring relative reliance on one's own abilities as well as on the AI system's capabilities. 
Studies employ the objective of building reliance by capturing the correctness of AI advice as well as the complementary task performance. 
We also briefly provide summaries of study designs \changed{and how interventions are used} to operationalize appropriate reliance constructs. 
Research increasingly highlights the need to gain an understanding of appropriate reliance on AI advice~\cite{schemmer2023appropriate, cabitza2023ai, lai2023towards, dellermann2019hybrid} to improve human-AI decision-making scenarios.
Still, to the best of our knowledge, limited work has evaluated the nascent field of exploring objective metrics used to study humans' appropriate reliance in human-AI decision-making research.
The main contributions our work makes are as follows:

\begin{enumerate}
    \item An analysis of recent research (up to 2025) that studies human-AI appropriate reliance empirically. 
    \item An argument on the operationalization of metrics for measuring human-AI appropriate reliance in contrast to metrics for measuring mere reliance.
    \item A synthesis of current gaps and recommendations for future work in human-AI appropriate reliance. 
\end{enumerate}

\section{Background and Related Work}
\label{sec-background}
Trust is fundamental to determining humans' willingness to adopt automated systems~\cite{hoff2015trust}.
To understand trust, theoretical frameworks from psychology are used in various studies~\cite{lee2004trust, hancock2023and, baier1986trust, tversky1981framing, tversky1989rational}.
The concept of trust has evolved over the years, from trust in automation (TiA)~\cite{lee2004trust, hoff2015trust} to trust in AI-based systems~\cite{choung2023trust, patricia2024understanding, leichtmann2023effects, vereschak2021how}.
By definition, trust is defined as an attitude that a system will help its users achieve in a situation characterized by uncertainty~\cite{lee2004trust}.
Commonly, trust can be evaluated as \textit{perceived, demonstrated, and actual} trust~\cite{wischnewski2023measuring, mehrotra2024systematic}. 
The \textbf{perceived} trust is based on users' perceptions or beliefs about systems~\cite{hoff2015trust, baier1986trust, lee2004trust}.
Perceived trust can give an inaccurate measure because the actual trustworthiness of the system may not be correctly perceived by the users~\cite{mehrotra2024systematic}.
The \textbf{demonstrated} trust accounts for the system's functionality, i.e., how trustworthy a system is in achieving the required performance under specific situations. 
\textbf{Actual} trust or trustworthiness of the system is more complex, which needs to capture \textit{why and how} people trust systems~\cite{schlicker2021towards}.
The actual trust may not be measured accurately, as it can be hindered by various factors (e.g., interfaces that do not present systems in their full capacity)~\cite{wischnewski2023measuring}. 

\subsection{From Trust to Reliance}
AI-assisted decision-making research operationalizes \textit{trust} and \textit{reliance} measures, where trust generally reflects the users' subjective perception (i.e., affective or emotional attitude) towards others or systems~\cite{lee2004trust, mehrotra2024systematic}, while reliance pertains to their (observable) objective behavior~\cite{scharowski2023exploring, dzindolet2003role, hoff2015trust}.
Trust captures a broader impression/demonstration of an AI system for its users while reliance covers a comprehensive case-by-case discrimination~\cite{schemmer2023towards, scharowski2022trust}.
\changed{For example, reliance measurements facilitate understanding whether users adopt AI assistance by accepting or rejecting AI advice in their decision-making~\cite{schemmer2023appropriate}.}
Reliance can also be related to demonstrated trust \changed{(i.e., adopting AI advice due to its trustworthiness)}, which signifies an evaluation of behavioral aspects, as used in measuring trust on AI~\cite{schemmer2023appropriate, eckhardt2024survey}.
The assessment of demonstrated trust can be viewed as a manifestation of user behavior (e.g., agreement or aversion with AI)~\cite{wischnewski2023measuring}. 
Initial works~\cite{hoff2015trust} mostly studied users' trust; however, trust measures do not necessarily enhance reliance~\cite{schemmer2023towards}, measurement of which has gained attention in recent work~\cite{lai2023towards, eckhardt2024survey}.

Definitions of trust or reliance can be contested in terms of their usefulness to clearly differentiate humans' abilities to rely on the AI advice~\cite{wischnewski2023measuring}. 
For example, one might, without assessing AI advice appropriately, put a high trust in a system that aligns with them. 
In such cases, an increase in trust can lead to an increase in reliance (or vice versa) if the system is highly accurate, though it may or may not always be correct~\cite{lee2004trust, schemmer2023towards}.
Oppositely, one might not trust a system highly but may have to rely on it~\cite{chen2023understanding}.
Hence, such behavior of trust and reliance is not substantiated by the actual case-to-case performance of the system and the human~\cite{wischnewski2023measuring, mehrotra2024systematic}. 
For instance, capturing this behavior does not inform whether the reliance was formed due to an overlap of the human and the AI's systems, i.e., when they agree with correct and incorrect AI suggestions or otherwise.
\changed{Therefore, it is important to consider reliance separately as a behavioral construct to understand how humans adopt AI advice in their decision-making and whether such adoption is appropriate, i.e., correct reliance on AI advice~\cite{cabitza2023ai, vasconcelos2023explanations}.}

\subsection{Human-AI Appropriate Reliance}
\changed{Understanding humans' reliance behavior with AI requires evaluating case-to-case performance in decision-making scenarios.}
To understand such behavior, research explores an approach called \textit{Judge-Advisor-System (JAS)}~\cite{sniezek1995cueing}, where a human acts as judge and gets decision assistance from AI.
For building reliance, this approach has to solve two main challenges: 1) the advice/advisor should be credible, and 2) humans should trust and then rely on the advice/advisor~\cite{lai2023towards}. 
\changed{While addressing these challenges, users may be able to build reliance on AI, but it cannot necessarily always be appropriate, i.e., when AI provides incorrect advice presented as correct, and users accept this advice without being able to recognize it as incorrect~\cite{schemmer2023appropriate}.}
Therefore, a more granular definition of appropriate reliance is defined~\cite{schemmer2023appropriate} as \textit{``humans' ability to discriminate correct and incorrect AI advice and to act upon that discrimination.''}
In addition, it is argued that users should neither \textit{over-rely} nor \textit{under-rely} on the AI advice~\cite{schemmer2023towards}. 
\textit{Over-reliance} is when users accept incorrect AI advice, and \textit{under-reliance} is when users reject correct AI advice, \changed{failing to achieve complementary performance.}
Hence, the appropriate reliance entails that humans can differentiate between correct and incorrect AI recommendations, and prevent over-and under-reliance. 
In this work, we investigate how research measures users' appropriate reliance, over-reliance, and under-reliance on AI advice.

Assessing users' over- or under-reliance may require an understanding of several underlying factors for measurement.
For instance, users may over-rely on the system's advice due to their belief in the system's superiority~\cite{jennifer2019algorithm, kahneman2009conditions, parasuraman2010complacency, wiener1981complacency}, initiated by \textit{complacency} or \textit{automation bias}.
Conversely, users may refrain from using the AI system due to \textit{performance bias}, perceptions of \textit{self-efficacy} (the ability to make decisions without advice), or purposefully considering the system's advice as inferior, demonstrating \textit{algorithmic aversion}~\cite{sutton2022extension, dietvorst2015algorithm}.
For example, if a system has higher accuracy, users may rely on all the decisions it suggests and achieve higher performance or greater reliance on AI; however, this is not appropriate, as users are merely complying with the system.
\changed{Such behavior may also lead to several biases, including automation bias~\cite{romeo2026exploring}, leading to inappropriate reliance on AI systems in decision-making.}
At the same time, if we only use the alignment with AI as a metric to estimate reliance, we can say users can (better) rely on the AI system. 
Still, it does not tell us about their behavior, for instance, when they were correct (or incorrect) and how AI advice influenced their behavior~\cite{schemmer2023appropriate}.

\subsection{Related Research Reviews}
Several existing reviews have explored human-AI collaboration research~\cite{bertrand2022how, wischnewski2023measuring, mehrotra2024systematic, eckhardt2024survey, lai2023towards, romeo2026exploring, raees2026people}.
Bertrand et al.~\cite{bertrand2022how} analyzed how explanations affect decision-making, examining users' cognitive biases and trust.
\changed{Romeo et al.~\cite{romeo2026exploring} reviewed research on \textit{``automation bias''} influencing humans' trust and reliance in human-AI collaboration, highlighting the impact of factors such as AI literacy, expertise, and explanations. 
Automation bias is a significant factor leading users to build trust in AI and align with AI advice~\cite{romeo2026exploring}.}
Wischnewski et al.~\cite{wischnewski2023measuring} evaluated studies on the calibration of trust in automated systems up to 2022.
This work~\cite{wischnewski2023measuring} analyzes calibration and metrics for trust in autonomous and robotic systems.
In this work~\cite{wischnewski2023measuring}, the authors distinguish between the perspectives of measuring perceived and demonstrated trust on AI systems.
In a similar context, Mehrotra et al.~\cite{mehrotra2024systematic} surveyed research from 2012-2022, synthesizing several perspectives on forming trust with automation and AI systems.
Their work examines historical perspectives on trust and explores existing disagreements over definitions and measures.

In human-AI decision-making research, Lai et al.~\cite{lai2023towards} analyzed empirical studies in human-AI complementarity, evaluating literature in human-AI decision-making up to 2021.
Eckhardt et al.~\cite{eckhardt2024survey} surveyed human-AI reliance research by employing a socio-technical lens. 
Raees et al.~\cite{raees2026people} conducted an analytical review to understand how people rely on AI, analyzing the characteristics of people (i.e., participants) and tasks used in the experiments. 
Reviews~\cite{lai2023towards, eckhardt2024survey, raees2026people} differentiate the objective metrics for reliance from the subjective trust, and critique the use of subjective metrics for measuring reliance. 
While these most up-to-date reviews highlight the needs and contexts for understanding trust and reliance, there is limited analysis of research that examines the metrics of human-AI appropriate reliance in detail. 
\changed{For instance, Eckhardt et al.~\cite{eckhardt2024survey} surveyed research, discussing constructs in human-AI \textit{reliance}, considering literature up to 2023. 
While their survey~\cite{eckhardt2024survey} highlights the need to distinguish reliance from appropriate reliance, it contains limited work that particularly focuses on appropriate reliance.
In addition, several studies exploring appropriate reliance have been published more recently.}
Hence, an in-depth synthesis and evaluation of the objective metrics used for assessing appropriate reliance is essential to form a consensus and inform future research, along with extending existing research~\cite{raees2026trust}. 
Due to a recent growth in human-AI reliance studies~\cite{eckhardt2024survey}, it is important to build upon the existing knowledge to augment the discussion on human-AI appropriate reliance in research. 

\textbf{Our Contribution.}
Despite recent research emphasizing the need to differentiate subjective trust from objective reliance, limited work has focused on evaluating the relevant assessment methods used to study human-AI appropriate reliance~\cite{lai2023towards, schemmer2023appropriate, eckhardt2024survey}.
\changed{The distinction between measurements and the constructs for reliance and appropriate reliance has been overlooked in recent work~\cite{eckhardt2024survey, lai2023towards}.}
As research increasingly focuses on assessing users' reliance on AI advice using objective metrics~\cite{schemmer2023appropriate, cabitza2023ai}, it is important to analytically evaluate recent studies exploring appropriate reliance.
In this work, we use a standardized method to analyze studies by specifically focusing on research in human-AI \textbf{appropriate reliance} and covering recent work.
We distinguished between definitions of trust, reliance, and appropriate reliance, which we believe need to be explicit to advance the current research in this domain.
Our work aims to add conceptual clarity on human-AI appropriate reliance research, evaluating different views and metrics for measurements.

\section{Methodology}
\label{sec-method}
\begin{figure*}
  \centering
  \includegraphics[width=\linewidth]{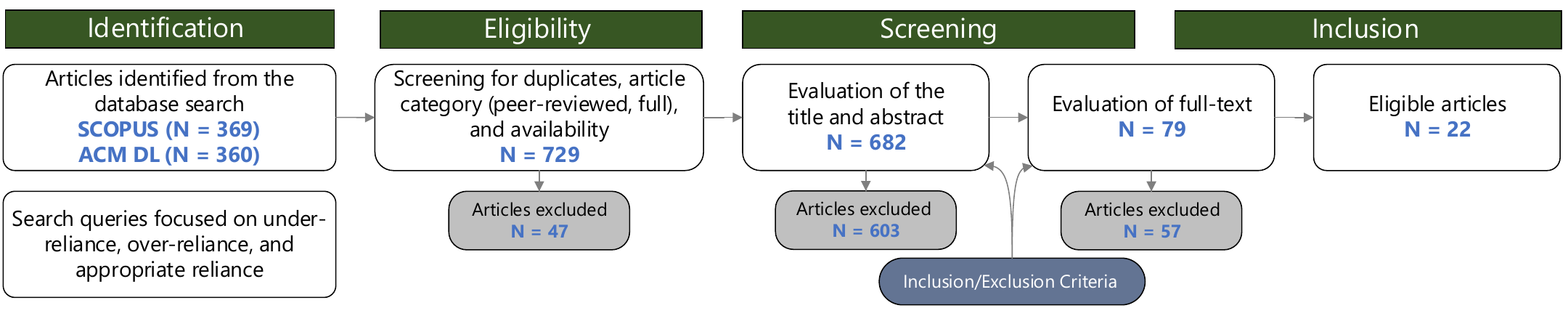}
  \caption{The PRISMA~\cite{page2021prisma} framework was used to assess research studies. The primary search was conducted on SCOPUS, with a secondary search on the ACM Digital Library (ACM DL).}
  \Description{Figure 1. The figure shows an overview of the prisma framework, a step-by-step process to collect, process, and analyze research documents. Studies are collected from databases, with 729 in total. Then, the inclusion and exclusion criteria removed 47 duplicates and non-peer-reviewed articles, 611 studies by title and abstract screening, and 49 by full text review, leaving 22 studies.}
  \label{fig-prisma}
\end{figure*}

We followed a systematic protocol (i.e., the PRISMA~\cite{page2021prisma} framework, as summarized in Figure~\ref{fig-prisma}) to identify and analyze literature, as adopted in prior studies~\cite{raees2024explainable, snyder2019literature}.
First, we searched the ACM Digital Library (DL) with keywords on ``human-AI reliance'' and ``human-AI decision-making''. 
We selectively skimmed through several articles matching the search terms.
Based on the analysis of identified articles, we devised keywords (\textit{``appropriate reliance'', ``over-reliance'', ``under-reliance''}) for a full search.
Keywords were complemented with \textit{``artificial intelligence'', ``machine learning'', and ``human-AI decision-making''} for focused context in searched databases.

The search was initially conducted on the SCOPUS database in mid-2025. 
Our initial analysis of SCOPUS results showed the prevalence of the topic studied at ACM venues, particularly within human-AI research. 
Hence, we conducted another search on the ACM DL in December 2025. 
In both full searches, the lower bound was set to 2018 to limit the results and focus on recent studies exploring the nascent topic of human-AI appropriate reliance. 
In total, we collected 729 search results across the queried databases. 
As we focused on evaluating the empirical studies of human-AI appropriate reliance, we applied the following inclusion and exclusion criteria as a funnel. 

\begin{enumerate}
    \item \textbf{Peer-Reviewed:}
    We only included full conference and journal papers. Short and workshop papers, books/chapters, and non-reviewed work (e.g., from arXiv) were excluded. In addition, secondary studies such as surveys and literature reviews were also excluded.

    \item \textbf{Human-AI Decision-Making:}
    From the remaining articles, we screened for studies with human subjects and human-AI decision-making contexts. We used title and abstract screening to remove articles without human subjects, decision-making context, and those that only explored theoretical aspects. 

    \item \textbf{Human-AI Appropriate Reliance Measurement with Objective Metrics:}
    We performed a full-text screening on the remaining articles, removing articles that did not evaluate 1) human-subject studies using quantitative methods, and 2) objective metrics to measure appropriate reliance (including over- and under-reliance).
\end{enumerate}

We applied the above-mentioned criteria sequentially and discussed for consensus among ourselves. 
After applying the criteria, we selected 22 papers that employed objective metrics of appropriate reliance as per the evaluation. 
\changed{Some studies also included subjective measures along with objective metrics, and those were not excluded.}
Several studies were excluded due to their focus only on metrics that did not capture appropriate reliance through objective metrics, i.e., \changed{only measuring reliance on AI using metrics like accuracy, agreements with AI, or switching decisions due to AI}. 
In another iteration, we looked at the relevant references of included studies to identify more papers using snowballing~\cite{wohlin2014guidelines} (by reviewing their titles and abstracts); however, no new results matching the criteria were found. 
The first author coded the papers across dimensions of evaluation metrics, experimental designs, and intervention methods used in studies. 
Afterwards, codes were collated to create groups and themes to report analysis and discussions. 

The selection of studies comes from top venues\footnote{CHI: Conference on Human Factors in Computing Systems. IUI: Conference on Intelligent User Interfaces. PACMHCI: Proceedings of the ACM on Human-Computer Interaction.} i.e., CHI, IUI, PACMHCI, and others.
Most selected studies were published between 2023 and 2025, showing the recent conceptualization of appropriate reliance.
The review protocol and coding are provided as an anonymized Open Science Framework (OSF) repository~\cite{osfstudylink2025} for the reader's evaluation.

\section{Analysis}
\label{sec-analysis}
In what follows, we provide our analysis, as coded in Table~\ref{tab-study-coding-short}.
Section~\ref{sec-4-1} examines the different views of reliance.
We analyze the metrics (objective and subjective) and measurements used in studies in Section~\ref{sec-4-2}.
Section~\ref{sec-4-3} discusses studies' intervention methods.
Finally, a concise analysis of the studies' experimental design elements is presented in Section~\ref{sec-4-4}.
The Appendices (Tables~\ref{tab-study-measures} -- \ref{tab-study-main-table}) provide summaries of evaluated research.

\begin{table*}
  \caption{Main dimensions for the analysis with three main views discussed in the literature to study appropriate reliance. The analysis also includes other dimensions, such as metrics, measurement protocols, and experimental design details.}
  \label{tab-study-coding-short}
  \Description{Table 2. A table providing an overview summary of dimensions of studies, with four columns: Dimensions, Categories, Sub-Categories, and Brief Explanations. The sub-categories column also shows the count of studies in each sub-category.}
  {\fontsize{8.5}{10}\selectfont
  \begin{tabular}{p{2.0cm} p{2.0cm} p{3.4cm} p{6.3cm}}
    \toprule
    Dimensions  & Categories  & Sub-Categories (\# Studies) & Explanation \\
    \midrule
      \multirow{3}{*}{\parbox{2.0cm}{Reliance Views}} & Traditional & - (10) & \multirow{3}{*}{\parbox{6.3cm}{\S \ref{sec-4-1} explores human-AI reliance views discussed in literature, differentiating from measures of trust.}} \\
      
       & Appropriateness & - (10) & ~ \\
                                              
        & Dominance & - (2) & ~ \\

        \midrule

        \multirow{4}{*}{\parbox{2.0cm}{Measuring \\Reliance on AI}} & \multirow{2}{*}{Metrics} & Objective Metrics (22) & \multirow{4}{*}{\parbox{6.3cm}{\S \ref{sec-4-2} examines objective and subjective metrics reported in the literature, along with their usage, in concurrent or multi-step decision-making protocols.}} \\ 
                                              
        & & Subjective Metrics (20) &  \\ \cline{2-3}
        
        & \multirow{2}{*}{\parbox{2.5cm}{Measurement \\ Protocol}} & Concurrent (4) & ~ \\
        
        & & Sequential (20)  & \\
        
        \midrule
           
        \multirow{3}{*}{\parbox{2.0cm}{Intervention \\Impact}} & \multirow{3}{*}{\parbox{2.0cm}{Appropriate \\Reliance}} & Positive (7) & \multirow{3}{*}{\parbox{6.3cm}{\S \ref{sec-4-3} explores the impact of common interventions on achieving appropriate reliance.}} \\
        
        & & Negative (5) & \\
        & ~ & No (Explicit) Effect (6) & ~ \\

        \midrule

        \multirow{7}{*}{\parbox{2.0cm}{Experimental \\ Design}}       & \multirow{3}{*}{Domains} & Education/Learning (7) & \multirow{7}{*}{\parbox{6.3cm}{\S \ref{sec-4-4} explores the contexts of applications (e.g., domain areas, implementation details), and participant expertise (e.g., background, domain knowledge), and the level of AI implementation (realism, complexity, models, etc.).}} \\
        &  & Business/Work-Context (7) & \\
        &  & Leisure/Sports/Arts (5) & \\ \cline{2-3}
        
        & \multirow{2}{*}{\parbox{2.0cm}{Participant \\Expertise}} & Domain/AI Users (2) & ~ \\
        & & Novices (20) & \\ \cline{2-3}

        & \multirow{2}{*}{AI Fidelity} & Actual (15) & ~ \\
        & & Simulated (7) &  \\                 

    \bottomrule
 \multicolumn{4}{l}{$^*$ Sub-categories (studies) are often non-exclusive.} 
\end{tabular}
}
\end{table*}

\subsection{Reliance Views and Concepts}
\label{sec-4-1}
Understanding appropriate reliance on AI systems as a research field has started growing recently, with various calls to differentiate it from understanding appropriate trust~\cite{cabitza2023ai, vasconcelos2023explanations, schemmer2023appropriate}.
Hence, we first examine how appropriate reliance is conceptualized in research among different views. 

In human-AI decision support, research has conceptualized the need to differentiate appropriate reliance as a behavioral construct and clearly examine its operationalization as distinctively as possible from the appropriate trust~\cite{schemmer2023appropriate, mehrotra2024systematic, eckhardt2024survey}.
Still, the conceptualization of appropriate reliance as a construct is in nascent stages and is often considered a mere adoption of AI advice. 
Hence, our first aim is to clarify the understanding of the different concepts surrounding appropriate reliance. 
In research, \textit{``Appropriate reliance''} is defined as a behavioral aspect with three patterns/views~\cite{schemmer2023towards, casolin2024towards} as \textbf{Traditional, Appropriateness, and Dominance}, which are expressed in Table~\ref{tab-xai-pattern}.
In all views, \changed{it is expected that the human decision-maker gets advice from AI and stays in control and is responsible for making the final decision as opposed to giving control to the AI system for final execution~\cite{wischnewski2023measuring}.}

The \textbf{traditional view}~\cite{vasconcelos2023explanations} assesses users' overall behavior (performance) with decision support for measuring over-reliance or under-reliance on AI. 
For instance, it captures the reliance behavior, arguing that users should follow correct AI recommendations and should not follow incorrect AI recommendations.
\changed{If users follow incorrect AI advice, they exhibit an over-reliance behavior, and if users do not follow correct advice, they show under-reliance behavior. }
This view is commonly studied without explicit definitions, as it mainly originates from concepts of demonstrated (or appropriate) trust, measured through users' objective behavior. 
Around 10 (45\%) studies in our corpus apply the traditional view for capturing users' reliance on AI advice, mostly being applied in earlier research studies.
This view focuses on the alignment between the perceived and actual performance of the system~\cite{mehrotra2024systematic, wischnewski2023measuring}.
The alignment is considered holistically instead of evaluating users' ability to differentiate decision-making outcomes, and measurements mainly focus on the accuracy of final decisions, \changed{i.e., adopting correct AI advice and rejecting incorrect AI advice.}
Though this view has objective metrics for over- and under-reliance, it does not fully drill down on the behavior users depict during the assessment~\cite{vasconcelos2023explanations, yang2020how, buccinca2021trust}.
For instance, this view does not account for whether users were able to identify errors in AI advice or in their own judgment and then make corrections to their decisions. 
 
Schemmer et al.~\cite{schemmer2023appropriate} define the \textbf{appropriateness view} as attempting to understand the users' relative self-reliance (RSR) and relative AI reliance (RAIR).
Relative self-reliance is a metric that describes the situation in which humans correctly reject wrong AI advice and make the correct decision independently, i.e., without being influenced by the AI's wrong recommendation.
Relative AI reliance captures the user's shift to a correct AI advice, i.e., the situation in which the user initially makes an incorrect decision but subsequently changes it to align with the correct AI recommendation.
These metrics capture the user's ability to rely on the AI advice during decision-making more appropriately~\cite{eckhardt2024survey}. 
For instance, in cases where users' decisions are different than systems' decisions (incorrect/correct), it becomes important to assess whether users can discriminate between such advice, i.e., can they contradict and make correct decisions using self or relative reliance.
Schemmer et al.~\cite{schemmer2023appropriate} argue that the metrics in the traditional view can have a major disadvantage of not knowing whether the reliance on AI advice stems from a correct discrimination of the AI decisions by humans or simply by an overlap of human and AI decisions.
As this view was introduced in 2023~\cite{schemmer2023appropriate}, 10 recent studies (45\%) apply it for measuring users' reliance on AI advice.  

\begin{table*}
  \caption{Views exploring trust and reliance in research. The Judge-Advisor System (JAS) pattern makes users decide before and after AI assistance. 1: correct advice/decision or 0: incorrect advice/decision w.r.t. the ground truth. In AI-assisted decision-making, measuring appropriate reliance has become an important challenge.}
  \label{tab-xai-pattern}
  \Description{Table 2. A table explaining different views of reliance. The first three columns represent human decisions, AI advice, and the final decision for correctness. The remaining columns show three views: naming appropriateness view, traditional view, and dominance view.}
  {\fontsize{8}{10}\selectfont
  \begin{tabular}{p{1.3cm} p{1.3cm} p{1.3cm} p{2.9cm} p{2.9cm} p{3.3cm}}
    \toprule
    \centering Human Decision &  \centering AI Advice &  \centering Final Decision & Traditional View~\cite{vasconcelos2023explanations} & Appropriateness View~\cite{schemmer2023appropriate} & Dominance View~\cite{cabitza2023ai} \\
    \midrule
    \centering 0& \centering 0& \centering 0&Over-Reliance& N/A &Detrimental Reliance \\
    \centering 0& \centering 0& \centering 1&Appropriate Reliance& N/A &Beneficial Under-Reliance \\
    \centering 0& \centering 1& \centering 0&Under-Reliance&Incorrect Self-Reliance&Detrimental Self-Reliance \\
    \centering 0& \centering 1& \centering 1&Appropriate Reliance&Correct AI Reliance&Beneficial Over-reliance \\
    \centering 1& \centering 0& \centering 0&Over-Reliance&Incorrect AI Reliance&Detrimental Over-Reliance \\
    \centering 1& \centering 0& \centering 1&Appropriate Reliance&Correct Self-Reliance&Beneficial Self-Reliance \\
    \centering 1& \centering 1& \centering 0&Under-Reliance& N/A &Detrimental Under-Reliance \\
    \centering 1& \centering 1& \centering 1&Appropriate Reliance& N/A &Beneficial Reliance \\
    \bottomrule
\end{tabular}
}
\end{table*}

Another important but slightly less applied view is defined by Cabitza et al.~\cite{cabitza2023ai}, called the \textbf{dominance view}.
The dominance view captures the dominance that technology exerts on users, which can be detrimental or beneficial in overall decision-making~\cite{sutton2022extension}. 
This view aims to capture how exertion by technology (AI) can influence users' appropriate reliance and is often studied in high-stakes domains to understand the beneficial and detrimental effects of using AI advice, \changed{as well as biases induced by the way AI advice is presented.} 
The dominating influence of AI can make users take a more subservient role (\changed{leading to confirmation bias or automation complacency}), and defer their decision-making to the technology (AI). 
When the influence of the technology is positive, i.e., such that the AI helps users avoid mistakes or make better decisions they would have made without getting AI advice, it will positively improve the overall (beneficial) reliance on the AI advice. 
However, the dominance can also negatively mislead users to make more mistakes, or over-rely on the technology (AI)~\cite{cabitza2023ai}, resulting in detrimental reliance.

Extensions of the discussed views are also explored, for instance, Morrison et al.~\cite{morrison2024impact} expand the appropriateness view from Schemmer et al.~\cite{schemmer2023appropriate}, including the dimension of incorrect explanations, i.e., where AI predictions can be correct, but their explanations can be incorrect or vice versa.
The concept of imperfect explanations~\cite{morrison2024impact} can exist regardless of the correctness of AI systems. 
For instance, explanations of AI systems may oversimplify or incorrectly portray the underlying decisions, leading to under- or over-reliance on the AI advice. 
Besides, explanations of AI systems are only one of the many components that impact reliance measurement, as explained in the following sections. 
Some studies also use other constructs that may partially conform with appropriate reliance, such as user agreeableness, or switching decisions with AI advice~\cite{buccinca2021trust, buccinca2025contrastive}, considering the correctness of final decisions.

\custombox{\textbf{Key insights.} 
The literature mainly explores three main views to assess users' appropriate reliance on AI advice. The \textit{traditional view} uses over-reliance and under-reliance measures based on overall behavioral patterns. The \textit{appropriateness view} uses relative AI reliance (RAIR) and relative self-reliance (RSR)  to measure users' independent and AI-dependent decisions. The \textit{dominance view} measures the benevolent or detrimental exertion of technology for final decisions.
Studies have also used extensions of these views to understand users' appropriate reliance on AI advice, but the research on consensus and validity of these views across domains remains limited. 
Defining clarity on measuring appropriate reliance is highly important, as highlighted in related studies of appropriateness and dominance~\cite{schemmer2023appropriate, cabitza2023ai}.
}

\subsection{Metrics and Measurements}
\label{sec-4-2}
\begin{figure*}
  \centering
  \includegraphics[width=\linewidth]{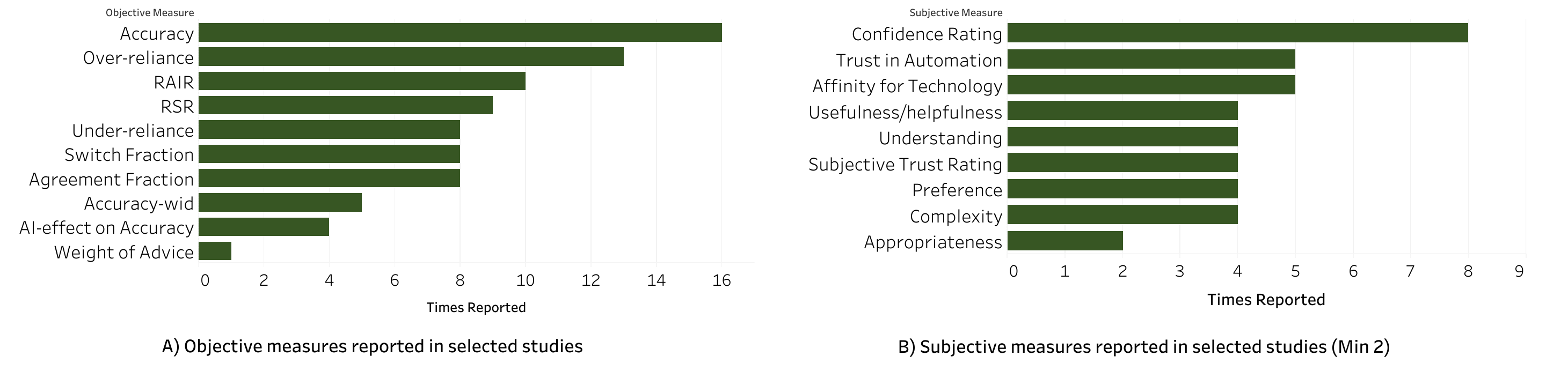}
  \caption{\textbf{Left}: The prevalence of objective metrics reported in selected studies. Accuracy is the most widely used measure. Common metrics for appropriate reliance used are over-reliance, under-reliance, RAIR, RST, and Accuracy-Wid. Studies also measure users' reliance on AI through agreements or switches.  \textbf{Right}: The prevalence of subjective metrics reported in selected studies. Studies commonly use confidence, trust, usefulness, and understanding ratings to measure subjective experiences. Some studies also capture users' complexity, preferences, and appropriateness to understand experiences better.}
  \Description{Figure 5. The figure shows two bar graphs to show the occurrence of objective (on the left) and subjective (on the right). On the left graph, accuracy has 16 counts, followed by over-reliance, 13, RAIR, 10, and RSR, 9. The under-reliance, agreement fraction, and switch fraction all have 8. Accuracy-wid has 5, AI-effect on accuracy has 4, while WOA has 1. On the right graph, confidence has 8, subjective trust and usefulness have 4, understanding has 4, trust in automation has 5, affinity in technology has 5, mental demand, appropriateness, and complexity have 4.}
  \label{fig-measures}
\end{figure*}

Here, we evaluate the metrics and measurements that studies have employed.
Figure~\ref{fig-measures} shows the prevalence of both objective and subjective metrics reported in studies.

\subsubsection{Reliance Metrics}
\changed{Studies use several metrics to measure users' adoption of AI advice to assess their reliance on AI.}
In selected studies, \textbf{decision accuracy}~\cite{cau2023effects} is a commonly used (16 studies) metric that measures the correctness of final decisions. 
However, it does not directly distinguish whether reliance was appropriate at a granular level, and only evaluates whether the final decision is correct or not.
While studies also employ other metrics that aim to capture users' behavior, they do not directly report to assess their appropriate reliance. 
For instance, \textbf{agreement fraction} (8 studies)~\cite{he2023knowing} and \textbf{switch fraction} (8 studies)~\cite{he2025conversational} are often used together to see how users agree or shift their decisions with AI, respectively. 
Agreement fraction measures how frequently (or what percentage of decisions) users agree with AI advice, and switch fraction measures how frequently (or what percentage of decisions) users change their initial decision after seeing the AI advice. 
Neither of these metrics, however, captures whether the reliance was appropriate, i.e., the correctness of user decisions in cases when AI advice was correct or incorrect.
Along with accuracy, around 5 studies also measure AI's effect on accuracy~\cite{buccinca2021trust} as a means to test if providing AI advice (against no advice) can improve users' reliance. 

\subsubsection{Appropriate Reliance Metrics}
\changed{Existing reviews~\cite{eckhardt2024survey, lai2023towards} evaluated metrics for overall reliance and have overlooked their distinction from metrics needed for capturing appropriate reliance explicitly. }
Recent studies explore the metrics for capturing appropriate reliance by applying traditional metrics like \textbf{over-reliance}~\cite{buccinca2021trust} (13 studies) and \textbf{under-reliance}~\cite{cau2023effects}  (8 studies) as separate metrics, which capture when users blindly agree with incorrect advice and neglect correct advice, respectively.
In terms of metrics aligned with the appropriateness~\cite{schemmer2023appropriate} view, studies also use \textbf{relative AI reliance} (RAIR) (10 studies) and \textbf{relative self-reliance} (RSR) (9 studies), as defined in Section~\ref{sec-4-1}.
Likewise, Accuracy-wid~\cite{he2023stated} (accuracy with initial disagreement -- 5 studies) is used to capture users' appropriate reliance when they disagree with the AI advice initially. 
Cabitza et al.~\cite{cabitza2023ai} also use \textit{weight of advice (WOA)} as a collective metric to measure if the users change their decision due to AI's assistance or exertion towards a correct decision.
Still, most studies rely on fixed reliance metrics (i.e., for each interaction), and user reliance development over time is under-explored~\cite{swaroop2025personalising}. 
The definitions of the above-mentioned metrics, along with others, are provided in Appendix~\ref{app-metrics-defs}.

\subsubsection{Measurement and Evaluation}
The point at which AI advice is given, i.e., using a \textit{concurrent or sequential} (multi-step) protocol, also affects the choice of metrics as both protocols slightly differ in the way they use reliance metrics~\cite{schemmer2023towards}.
For instance, concurrent approaches lack switching or a change of decision after taking the advice, hence, measuring fewer objective metrics.
Therefore, 90\% of the studies (20) in our corpus mainly use a two-step (or multi-step) process where humans first make the decisions, investigating the task themselves, and then are provided with the AI advice, subsequently having the option to revise their decision. 
Though not very common, around around 18\% of the studies (4 studies), often as part of multi-study experiments, use concurrent process.
\changed{The use of two-stage decision-making has become a standard in recent studies, as an initial human decision is essential to capture for the employed metrics for capturing appropriate reliance.}
Some cases include reliance development over multi-stage decisions, where users explore additional information in intermediary steps before making decisions.
Around 90\% (20) of the studies in our corpus rely on assessing and reporting user interaction on these metrics using \textbf{empirical analysis}, while a few (2 studies) use \textbf{mixed methods}, showing the growing prevalence of objective measurements \changed{compared to evaluations reported in existing reviews~\cite{eckhardt2024survey, lai2023towards}}.

\subsubsection{Subjective Metrics}
Studies often supplement the objective metrics of human reliance on AI with various \textbf{subjective metrics}, aiming at capturing user perceptions towards AI. 
Commonly used subjective metrics include the users' \textbf{confidence}~\cite{yang2020how} in their decisions (8 studies), Trust in Automation~\cite{he2025conversational} (5 studies) or perceived \textbf{trust}~\cite{he2023knowing} (4 studies), \textbf{mental demand}~\cite{buccinca2021trust} (5 studies), \textbf{usefulness}~\cite{ma2025towards} (4 studies), \textbf{understanding}~\cite{chiang2023two} of advice (4 studies), and others~\cite{cau2025influence, ma2024you, swaroop2025personalising}.
Subjective metrics focus on measuring the perceived usefulness and understanding of AI advice to capture users' experiences with the system. 
These metrics are mostly self-assessed attributes and are also used to understand the impact on user performance, but they do not necessarily help measure actual reliance behavior. 
Survey feedback is also used to provide deeper insights about human behavior in reliance strategies. 

A summary of the objective and subjective metrics used by the examined literature is provided in Table~\ref{tab-study-measures} (Appendix).
Due to the variance in using metrics to measure users' performance, there is still limited consensus on common measurements of humans' appropriate reliance on AI across the studies.
The combination of objective and subjective metrics does not necessarily lead to better predictions of reliance; for instance, users can express high trust towards a system that aligns with them, without any objective indication of appropriate reliance.

\custombox{\textbf{Key insights.} 
The literature reports the use of a variety of objective and subjective metrics to evaluate the users' reliance and appropriate reliance, as well as their perceptions towards AI. 
In terms of \textbf{objective metrics}, we find that over-reliance, RAIR, RSR, and under-reliance are commonly used as appropriate reliance metrics. 
Other metrics, such as switch fractions and agreement fractions, are used, but for capturing user behavior without the focus on appropriate reliance. 
The accuracy metric is used in both cases.
In terms of \textbf{subjective metrics}, we find several metrics that only measure perception or feedback, for instance, to capture trust and confidence for gauging the users' perception of AI systems.
However, subjective impressions (such as trust and confidence) do not necessarily reveal whether users can build an appropriate reliance on AI. 
}

\subsection{Interventions for Human-AI Reliance}
\label{sec-4-3}
Here, we summarize the most commonly applied interventions and their impact on achieving appropriate reliance.
Overall, studies show mixed results regarding the improvement of reliance on AI (see Table~\ref{tab-study-effects}), and cite limitations in forming appropriate human-AI reliance.

\subsubsection{\changed{Engagement and Effort}}
Studies have used several interventions (e.g., cognitive forcing~\cite{buccinca2021trust} or frictional AI~\cite{cabitza2023ai}) with metrics to capture users' over-reliance. 
Metrics also measure how users engage with AI systems~\cite{buccinca2021trust, swaroop2025personalising} to capture whether they can appropriately assess the AI advice or not. 
For instance, if users can analytically engage with AI advice and spend time understanding it, they may be able to reduce their over- or under-reliance on AI~\cite{vasconcelos2023explanations}. 
Explanations of AI systems help users in understanding AI advice, but they have often been reported to cause over-reliance~\cite{buccinca2021trust, swaroop2025personalising}. 
In practice, people weigh the potential benefits of cognitive effort against its perceived cost~\cite{kool2018mental}, a phenomenon also referred to as \textit{cognitive-effort-discounting}~\cite{westbrook2013subjective}. 
For example, people are less likely to over-rely on the AI system's advice when the benefit of obtaining the correct answer substantially outweighs the risk of over-reliance (e.g., avoiding sending an unprofessional email to a boss), compared with situations in which the benefit-risk tradeoff is low (e.g., assessing the sentiment of a tweet)~\cite{vasconcelos2023explanations}.
Hence, people's reliance on the AI is also impacted by the value of the decision and the effort it takes to make correct decisions~\cite{salimzadeh2024dealing}, which needs to be measured. 

\subsubsection{Task Complexity}
Task difficulty is often not considered in measuring humans' appropriate reliance on AI advice~\cite{salimzadeh2023missing}; however, it can significantly impact user behavior with AI.
Hence, the same metrics may not be applicable if the tasks have different complexity and users' stakes. 
For instance, task complexity increases information overload pertaining to the user's perception and capabilities~\cite{parkes2017effect}, which also should be taken into account for measuring reliance.
Some tasks can be assigned to rely on AI advice easily (easy or low-stakes), while others may require more consideration~\cite{swaroop2024accuracy, swaroop2025personalising}.
Therefore, considering the task complexity for measuring appropriate reliance is an important area of research to explore~\cite{salimzadeh2023missing}.
For instance, when users lack expertise in a task, they tend to (inappropriately) rely more heavily on AI advice, especially as the task becomes more complex~\cite{das2020leveraging, salimzadeh2024dealing}.
\changed{Comparing whether existing metrics produce similar results in different task settings for a similar population can inform their validity for measuring appropriate reliance across settings~\cite{salimzadeh2023missing}.}
Research suggests that when the task is difficult, the design should help users make better decisions through reflections~\cite{sengers2005reflective, kahneman2011thinking}. 
Hence, assessing or measuring whether users reflect in their decision-making can provide more insights into their reliance behavior~\cite{buccinca2021trust}.

\begin{table*}
\caption{Common interventions and their influence on achieving appropriate reliance under the experimental conditions. Positive influence means that the respective methods were found useful in improving users' behavior for appropriate reliance.}
  \label{tab-study-effects}
  \Description{Table 4. A table showing intervention effects, with three columns: Objective, Impact, Intervention, and Study.}
  \begin{tabular}{ll}
  \toprule
    Interventions & Influence \\
\midrule

    Cognitive Forcing~\cite{buccinca2021trust}, Deductive-XAI~\cite{cau2023effects}, Second Opinion~\cite{chiang2023two, lu2024does}, Cognitive Effort~\cite{vasconcelos2023explanations} & Positive  \\
    
    \midrule
    
   Task Difficulty~\cite{vasconcelos2023explanations, salimzadeh2024dealing}, Poor Impression~\cite{he2024err}, Self Assessment~\cite{he2023knowing}, Domain Knowledge~\cite{morrison2024impact} & Negative   \\
   
   \midrule
    
    Example-based XAI~\cite{buccinca2021trust, schemmer2023towards, cau2025influence}, Feature-based XAI~\cite{schoefer2024explanations, cau2025influence}, Showing AI Uncertainty~\cite{ma2024you} & No Effect \\

          \bottomrule
\end{tabular}
\end{table*}

\subsubsection{User Expertise}
Users' expertise or skills with AI systems are also an important factor to measure~\cite{buccinca2025contrastive}, \changed{which is mostly captured as fixed subjective self-reports.}
Bu\c{c}inca et al.~\cite{buccinca2025contrastive} highlighted that measuring users' competence and skill can also inform reliance behavior, as AI support often fails to address the knowledge gaps that users seek to fill~\cite{buccinca2025contrastive}.
They~\cite{buccinca2025contrastive} showed that enhancing users' skills with AI systems is important for improving engagement and human decision-making. 
Similarly, measuring users' interaction with partial explanations can also be useful~\cite{lu2024does, ma2025towards}. 
Partial explanations align with a larger body of research on deliberation that encourages users to reflect on their decisions~\cite{ma2025towards}.
Morrison et al.~\cite{morrison2024impact} explored the metrics for capturing whether users can differentiate between correct and incorrect explanations of AI systems. 
Lu et al.~\cite{lu2024does} pointed out that active solicitation can also influence reliance, for instance, by getting a second opinion to form a consensus.
Different people trust AI differently~\cite{swaroop2025personalising}, and capturing their traits and real-time behavior can help define metrics that can be applied to study personality factors in human-AI appropriate reliance.
Capturing other demographics (e.g., education, literacy)~\cite{he2023knowing}, and more importantly, domain expertise are central to defining and evaluating accurate metrics for human-AI appropriate reliance. 

\custombox{\textbf{Key insights.} 
Certain methods, such as cognitive forcing and deliberation, show positive effects in building appropriate reliance. However, their operationalization in tasks is seldom measured in practice. 
In particular, assessing users' expertise (i.e., an estimation of their skills) and the difficulty of the task are important.
However, studies often do not relate user behavioral patterns (engagement, skills, expertise) and task complexity (easy to hard) with metrics used to capture appropriate reliance. 
Studies also indicate that several applied interventions have either a negative or, at best, a neutral effect on appropriate reliance. 
}

\subsection{Experimental Design}
\label{sec-4-4}

We also examine the experimental design of human-AI reliance studies.
Studies have employed various applications in various high- and low-stakes domains, as depicted in (Table~\ref{tab-study-domains}). 

\subsubsection{Study Participants}
Studies show the application of diverse tasks, but most (20 out of 22) of them use lay users or crowd workers as the study subjects performing the tasks. 
Such reliance on lay users and crowd-workers limits actual evaluations of AI systems with relevant stakeholders.
Studies recruit on average approximately 309 people (\textit{mean = 309.18, std = 273.93}). 
Around 50\% (11) of the examined studies acknowledge the limitations of recruiting laypeople and using tasks that are not relevant to their expertise or stakes. 
While evaluations with crowd workers are appealing when assessing target users, they only capture the general notion of human-AI decision-making~\cite{doshi2017towards}. 
Therefore, it may be possible that measurements in such experimental settings may not be realistic representations of actual context-specific scenarios. 
Many studies (45\%) did not use tasks that were relevant to the users' expertise.
It can also be argued that measurements may miss capturing users' reliance if they provide low-effort results~\cite{he2025conversational}.
Despite ensuring crowd workers can execute tasks~\cite{doshi2017towards}, it is also useful to employ reliance measurements with domain users for decision-making tasks.
Existing work~\cite{ramos2020interactive, dudley2018review} also highlights that running complex tasks with crowd workers may not be entirely representative and therefore may not report actual measurement of appropriate reliance on AI advice. 
Hence, further work is required to understand the assessment of metrics for tasks that are properly representative of decision-making scenarios explored in studies for comparative evaluations across contexts.

\subsubsection{Systems}
The systems or situations in which study experiments are conducted may also influence the metrics and results for human reliance on AI. 
Studies examine human decision-making using a mix of AI models (Table~\ref{tab-data-model}).
However, around 7 studies (33\%) do not construct real AI interventions, and, instead, use simulated~\cite{dahlback1993wizard} approaches, trying to replicate what users would believe or experience while interacting with AI-assisted decision support.
Although this is a well-accepted strategy to evaluate systems in research, non-realistic AI systems risk fully replicating a realistic AI and human behavior~\cite{swaroop2025personalising}.
The literature acknowledges that evaluating AI systems in their actual form through human subject experiments that involve realistic tasks can be important to assess reliance~\cite{doshi2017towards}.
Hence, using fictitious or non-representative tasks, especially when performed with crowd workers in low-stakes or irrelevant contexts (about the task at hand), may not represent real-world situations. 
Therefore, for an accurate assessment of metrics, the system implementations should also be uniform across studies. 

\custombox{\textbf{Key insights.} 
Studies have measured humans' appropriate reliance on AI in various application domains. However, the majority of the studies employed novice participants, and task scenarios may not be comparable across studies, thereby limiting the generality of the used metrics. 
The lack of uniform evaluations may considerably undermine the applicability of results from studies across tasks and participants' expertise. 
Many studies do not employ real AI systems; rather, they rely on simulated approaches, which further limit their operationalization and appropriate assessment across contexts.
}

\section{Discussion}
\label{sec-discuss}
Current research in human-AI decision-making focuses on assessing users' trust and reliance on AI systems through diverse metrics.
While existing work has focused on evaluating calibrated or appropriate trust using different metrics~\cite{mehrotra2024systematic, wischnewski2023measuring}, less research work has focused on measuring users' appropriate reliance on AI. 
Our analysis shows that there is a nascent but growing interest in empirical research to evaluate users' appropriate reliance on AI using objective metrics.
As highlighted in recent literature~\cite{cabitza2023ai, eckhardt2024survey}, there is a need for evaluations of metrics for measuring trust and reliance distinctively. 
For instance, by clarifying the constructs for reliance, metrics should aim to assess users' performance for case-to-case interactions and behavior, e.g., on being able to adopt AI advice to achieve a task.
\changed{We further argue that metrics for reliance and appropriate reliance also have a fine distinction, which needs to be more explicit in measurements and related constructs.}
Measuring objective performance on being able to discriminate correct/incorrect AI advice provides a realistic notion of users' appropriate reliance on AI advice, compared to using a mix of objective and subjective metrics~\cite{wischnewski2023measuring}. 
However, our analysis shows that the field is yet to settle on metrics to measure users' appropriate reliance on AI~\cite{lai2023towards, cabitza2023ai}, \changed{which is often mixed with measuring mere reliance~\cite{eckhardt2024survey}.} 

Investigating users' actual behavior with AI systems instead of only capturing their trust or agreement with AI is essential to test systems' appropriate use in realistic contexts. 
Our work conceptualizes and provides the current state of research on metrics for assessing appropriate reliance, highlighting the discord with measuring trust and reliance. 
Our analysis also distinguishes appropriateness of reliance from mere reliance as a conceptual framework.
Conceptually, it is important to distinguish between metrics for appropriate reliance, such as over-reliance, under-reliance, and relative reliance, from the existing metrics of accuracy, agreement, or switching, among others~\cite{lai2023towards, eckhardt2024survey}. 
Such a distinction will help clarify the need for more refined and standardized metrics for appropriate, over-, and under-reliance, as well as consolidate the research agenda in this domain. 
This consolidation can ensure that future empirical studies use standardized and validated metrics for measurement. 

\subsection{Potential Research Gaps}
In what follows, we briefly reflect on potential gaps and challenges we identified through our analysis of current human-AI appropriate reliance research.

\begin{itemize}

    \item Varying task complexity and users' expertise can elicit different reliance behavior, which needs to be measured across domains to validate metrics~\cite{salimzadeh2023missing}. Hence, studies show limitations in generalizing results across domains, as most metrics are not standardized~\cite{schoeffer2025ai}. In addition, intervention-specific behaviors are often under-substantiated. For example, studies introduce methods to enhance user engagement, deliberation, and other capacities to assess AI systems; however, they barely measure how users engage with such interventions. 

    \item Experiments with non-realistic tasks and a lack of users' expertise on those tasks may fail to assess users' reliance appropriately~\cite{gaube2021ai, veinot2018good}. Optimal measurements may come from studies that apply metrics in realistic evaluations instead of one-off experiments with crowd workers and proxy tasks~\cite{doshi2017towards}. Metrics may exhibit different outcomes when applied in domain-specific settings.

    \item \changed{Metrics to assess reliance and appropriate reliance have been mostly studied in classification-related scenarios where outputs have fixed ground truth and are considered binary (accept or reject AI advice). Other AI types may have different assumptions. For instance, generative AI outputs often do not have a fixed ground truth and may require different metrics, which are highly underexplored in research~\cite{hwang202580}.}
    
    \item Subjective metrics may only test users' perception of AI systems and do not validate their reliance. Their use reflects a lack of a common framework for selecting metrics using a standardized method~\cite{eckhardt2024survey}. 
    
    \item Broadly, studies focused on over-reliance and somewhat overlook the aspect of under-reliance, i.e., when users do not use the AI advice~\cite{swaroop2025personalising, swaroop2024accuracy}. Hence, equal consideration of both constructs is important for providing insights from both perspectives (over- and under-reliance) in the user studies.
    
    \item Research uses metrics that measure users' reliance on one-off interactions~\cite{schemmer2023towards, he2023knowing}. Capturing users' reliance in scenarios where system performance and users' reliance change over time is less explored~\cite{schoeffer2025ai, wang2023watch}. For instance, users' reliance may change while interacting with an AI system after a few iterations. Therefore, further research on such factors could provide empirical insights about \textit{progressive reliance} on AI advice. 

    \item Human-AI reliance research has under-explored the users' ability to manipulate, contest, or change the AI assistance, which can be an important factor in measuring their reliance~\cite{raees2024explainable}. Real AI systems may have the flexibility to adapt AI advice~\cite{dudley2018review}, and investigating reliance in setups where AI systems can be manipulated can induce more insights and elicit additional metrics for measurement.

\end{itemize}

\subsection{Recommendations for Further Studies}
Current research on human-AI reliance has several open questions that researchers can further explore.

\begin{itemize}
    \item Establishing definitions and terminologies for appropriate reliance is integral. We discussed three views explored in research: \textit{traditional}~\cite{vasconcelos2023explanations}, \textit{appropriateness}~\cite{schemmer2023appropriate}, and \textit{dominance} ~\cite{cabitza2023ai}. Although these views focus on objective metrics for measuring appropriate reliance on AI, there is a need to unify them as frameworks. \changed{Such frameworks could be identified based on the goal or utility of AI interventions. One such framework could unify the assessment of patterns when initial human decisions and AI advice differ in correctness. In practice, studies still use fragmented concepts and metrics of appropriate reliance.
    } 
    Standardizing the use of metrics for measuring appropriate reliance, as well as introducing extended metrics, is recommended~\cite{morrison2023evaluating}. 

    \item Over-reliance on AI can also lead users to lose their skills in the tasks, which should also be assessed by incorporating metrics for users' \textit{deskilling}~\cite{buccinca2025contrastive} and \textit{agency loss}~\cite{raees2024explainable}, \changed{which are rarely used in studies}. Assessing appropriate reliance is also important in situations where users are given the agency to contest and adjust an AI advice~\cite{dietvorst2015algorithm}. Further research can explore metrics for appropriate reliance in such settings. 
    
    \item Metrics for appropriate reliance can also be aligned to measure user engagement with AI~\cite{cornelissen2022reflection, miller2023explainable}. Engaging users with AI can help them contest, take ownership of their decisions, and establish critical engagement or mindful usage~\cite{hinrichs2024exploring}. Measuring how users take ownership of decisions while building appropriate reliance can inform more about their ability to evaluate AI. 

    \item \changed{Research also needs to explore metrics for appropriate reliance with generative AI~\cite{kim2024m, de2025cognitive}, as the mode of adopting AI advice transitions from binary to open choice~\cite{hwang202580}. Such outputs can have varying levels of correctness and value for different user groups, providing several opportunities for future research to explore.}

    \item Assessing users' \changed{domain expertise and self-confidence} about the task is essential for eliciting a realistic behavior~\cite{ma2023should}. \changed{For instance, measurements can enable capturing users' reliance on AI with superior performance to themselves, and self-reliance when AI underperforms~\cite{ma2024you}. However, due to over-reliance, users may not be able to assess the system's incorrect advice when the system underperforms.} Reliance on AI may also depend on \changed{\textit{situational/contextual factors}, \textit{cognitive traits}}, or \textit{individuality}~\cite{swaroop2025personalising, swaroop2024accuracy, de2025cognitive} for more realistic evaluations. 

    \item \changed{Understanding and modeling why inappropriate reliance happens (i.e., due to poor self-assessment, literacy, bias, or task stakes) can inform more realistic assessment metrics for appropriate reliance~\cite{guo2024decision}. Such understanding can inform intervention design to facilitate and measure appropriate reliance effectively.}
     
\end{itemize}

\subsection{Limitations}
We focused on research in human-AI appropriate reliance measurement, where the vast majority of studies have previously explored measuring trust.
This focus may have resulted in eliminating studies that use appropriate trust interchangeably with appropriate reliance. 
Our search was conducted using SCOPUS and the ACM Digital Library, similar to reviews in related domains~\cite{wischnewski2023measuring, lai2023towards}.
While our search was comprehensive, some work on human-AI reliance may exist outside the covered databases.
\changed{Though we searched on other engines (i.e., IEEE eXplore and Google Scholar), the results found beyond what were already in the dataset lacked exploring appropriate reliance constructs and objective metrics, and thus were not reported/included. 
Still, other specific libraries (e.g., AIS) can be explored to enhance search results.} 
In addition, despite careful evaluation, we acknowledge that results may be affected by screening bias and subjective interpretation during the application of inclusion and exclusion criteria. 

\section{Conclusion}
\label{sec-future}
We examined the recent literature on human-AI decision-making with a particular focus on research on humans' appropriate reliance on AI advice.
\changed{We first highlighted the distinction between metrics for measuring appropriate trust, reliance, and appropriate reliance.}
\changed{Building upon the definitions}, our analysis showed that there are three common views in research for conceptualizing reliance, namely, traditional, appropriateness, and dominance. 
We found that there are clear gaps in the constructs explored in the literature, as well as in measurements used in application contexts.
We discussed several open-ended research gaps and potential directions for future research to better understand and explore constructs for human-AI appropriate reliance.
We provide our study corpus, coding, and analysis as an anonymized but public resource~\cite{osfstudylink2025} for future work to expand upon.

\section{Generative AI Usage Statement}
No generative AI tools were used in the writing, editing, or preparation of this manuscript. All content was produced by the authors.

\bibliographystyle{ACM-Reference-Format}
\bibliography{refs}

\appendix

\section{Metrics}
\label{app-metrics-defs}

\subsection{Objective Metrics}

\[
\text{Over-reliance} = \frac{\text{Number of correct initial decisions changed to incorrect due to AI advice}}{\text{Total number of decisions with initial disagreement}}
\]

\[
\text{Under-reliance} = \frac{\text{Number of incorrect final decisions despite correct AI advice}}{\text{Total number of decisions with initial disagreement}}
\]

\[
\text{Relative AI Reliance (RAIR)} = \frac{\text{Positive AI reliance}}{\text{Positive AI reliance} + \text{Negative self-reliance}}
\]

\[
\text{Relative-Self-Reliance (RSR)} = \frac{\text{Positive self-reliance}}{\text{Positive self-reliance} + \text{Negative AI reliance}}
\]

\[
\text{Accuracy-wid} = \frac{\text{Number of correct final decisions with initial disagreement}}{\text{Total number of decisions with initial disagreement}}
\]

\[
\text{Weight of Advice (WOA)} = \frac{|\text{Advice} - \text{Initial estimate}|}{|\text{Final estimate} - \text{Initial estimate}|}
\]

\[
\text{Accuracy} = \frac{\text{Number of correct decisions}}{\text{Total number of decisions}}
\]

\[
\text{Agreement Fraction} = \frac{\text{Number of decisions same as the system}}{\text{Total number of decisions}}
\]

\[
\text{Switch Fraction} = \frac{\text{Number of decisions user switched to agree with the system}}{\text{Total number of decisions with initial disagreement}}
\]

\[
\text{AI-effect on Accuracy} = \frac{\text{Number of changes to correct decisions due to AI advice}}{\text{Number of correct AI decisions}}
\]

\subsection{Subjective Metrics}

\begin{itemize}
    \item \textbf{Subjective Trust Rating:} A self-reported measure of how much a user trusts the AI system.
    \item \textbf{Trust in Automation:} The belief that the AI system will perform tasks reliably and effectively.
    \item \textbf{Affinity for Technology:} A user's general attitude and comfort level with using technology.
    \item \textbf{Understanding:} The degree to which the user comprehends how the AI system works or makes decisions.
    \item \textbf{Confidence Rating:} A user's self-assessed certainty in their own decision or judgment.

    \item \textbf{Usefulness/Helpfulness:} Perceived benefit or support provided by the AI in completing the task.
    
    \item \textbf{Mental Demand / Task Load:} The cognitive effort required to complete the task with or without AI assistance.
    
    \item \textbf{Appropriateness:} How suitable or contextually correct the AI's behavior or involvement is perceived to be.

    \item \textbf{Comfort:} The ease and lack of stress experienced while interacting with the AI.
    
    \item \textbf{Preference:} A user's choice or inclination toward using the AI system over alternatives.
    
    \item \textbf{Complexity:} The perceived difficulty or intricacy of the task or the AI system itself.
    
    \item \textbf{Enjoyment:} The degree of pleasure or positive experience while using the AI.
    
    \item \textbf{Satisfaction:} Overall contentment with the interaction or outcome involving the AI.
    
    \item \textbf{Personality Traits:} Individual differences that may influence AI reliance behavior.
    
    \item \textbf{Motivation:} The internal drive or reasons behind engaging with the task or the AI system.
\end{itemize}

\section{Summaries of Review}

\begin{table}[htpb]
  \caption{Objective and subjective measures used in studies. The objective reliance measures are used to study how users fare with decisions. Trust is a commonly used subjective measure. Other measures are used to understand users' interaction with systems.}
  \label{tab-study-measures}
  \Description{Table 4. A table providing objective and subjective measures used in studies, with two columns: Category and Reported Measures. Objective reliance measures are used to study how users fare with decisions.}
  \begin{tabular}{@{}p{1.8cm} p{12.9cm}}
    \toprule
    Category & Reported Measures  \\
    \midrule
        
        Reliance & Accuracy~\cite{schemmer2023appropriate, he2023knowing, he2023stated, cau2023effects, chiang2023two, ma2024you, salimzadeh2024dealing, schoefer2024explanations, he2024err, swaroop2024accuracy, lu2024does, he2025conversational, swaroop2025personalising, buccinca2025contrastive, cau2025influence, ma2025towards}, Agreement Fraction~\cite{cabitza2023ai, he2023knowing, he2023stated, ma2024you, salimzadeh2024dealing, he2024err, he2025conversational, ma2025towards}, Switch Fraction~\cite{he2023knowing, he2023stated, ma2024you, salimzadeh2024dealing, he2024err, he2025conversational, ma2025towards}, AI-effect on Accuracy~\cite{buccinca2021trust, schemmer2023towards, cabitza2023ai, buccinca2025contrastive}, Weight of Advice~\cite{cabitza2023ai}, AI Safety~\cite{cabitza2023ai}, Human Error~\cite{buccinca2021trust} \\
        \midrule

        Appropriate Reliance & Over-reliance~\cite{yang2020how, buccinca2021trust, vasconcelos2023explanations, cau2023effects, chiang2023two, ma2024you, schoefer2024explanations, morrison2024impact, swaroop2024accuracy, lu2024does, swaroop2025personalising, cau2025influence, ma2025towards}, Under-reliance~\cite{yang2020how, cau2023effects, chiang2023two, ma2024you, schoefer2024explanations, morrison2024impact, lu2024does, ma2025towards}, RAIR~\cite{schemmer2023appropriate, schemmer2023towards, cabitza2023ai, he2023knowing, he2023stated, salimzadeh2024dealing, he2024err, morrison2024impact, he2025conversational, swaroop2025personalising}, RSR~\cite{schemmer2023appropriate, schemmer2023towards, cabitza2023ai, he2023knowing, he2023stated, salimzadeh2024dealing, he2024err, morrison2024impact, he2025conversational}, Accuracy-wid~\cite{he2023knowing, he2023stated, ma2024you, salimzadeh2024dealing, he2025conversational}, Deception of Reliance~\cite{morrison2024impact} \\
        \midrule

        Others & Subjective Trust Rating~\cite{yang2020how, buccinca2021trust, vasconcelos2023explanations, ma2025towards}, Trust in Automation~\cite{he2023knowing, he2023stated, salimzadeh2024dealing, he2024err, he2025conversational}, Affinity for Technology~\cite{he2023knowing, he2023stated, salimzadeh2024dealing, he2024err, he2025conversational}, Understanding~\cite{yang2020how, chiang2023two, he2025conversational, ma2025towards}, Confidence Rating~\cite{yang2020how, schemmer2023appropriate, cau2023effects, chiang2023two, ma2024you, lu2024does, cau2025influence, ma2025towards}, Usefulness/helpfulness~\cite{he2023knowing, he2023stated, swaroop2024accuracy, ma2025towards}, Mental Demand (and/or Task Load)~\cite{buccinca2021trust, ma2024you, he2024err, cau2025influence, ma2025towards}, Appropriateness~\cite{cau2023effects, buccinca2025contrastive}, Preference~\cite{yang2020how, buccinca2021trust, swaroop2025personalising, cau2025influence}, Complexity~\cite{buccinca2021trust, ma2024you, swaroop2024accuracy, ma2025towards}, Enjoyment~\cite{vasconcelos2023explanations, swaroop2025personalising, buccinca2025contrastive}, Satisfaction~\cite{ma2024you, ma2025towards}, Personality Traits~\cite{swaroop2024accuracy, swaroop2025personalising}, Motivation~\cite{swaroop2025personalising, buccinca2025contrastive}

        \\

    \bottomrule

\end{tabular}
\end{table}

\begin{table}[htpb]
  \caption{Research uses various types of decision-making tasks in several real-world and toy domains, ranging from high-stakes (healthcare) to low-stakes (leisure), to study human reliance patterns.}
  \label{tab-study-domains}
  \Description{Table 6. The table provides an overview of domains and decision tasks where the applications are used. The table has two columns labeled Domains and Decision-Task. }
  \begin{tabular}{@{}p{2.2cm} p{12.3cm}}
    \toprule
    Domains & Decision-Task  \\
    \midrule
        
        General & Hotel Reviews~\cite{schemmer2023appropriate, he2024err}, Hand-written Digits Identification~\cite{cau2023effects}, Leaf Classification~\cite{yang2020how}
        \\
        
        Critical Domains & Medical Diagnostic~\cite{cabitza2023ai}, Drug Prescription~\cite{swaroop2025personalising, swaroop2024accuracy}, Recidivism~\cite{chiang2023two}
        \\
        
        Business & Income Prediction~\cite{ma2024you}, Banking~\cite{brachman2022reliance}, Loan Approval~\cite{he2023stated}, Profession/Job Prediction~\cite{schoefer2024explanations, cau2025influence}
        \\
        
        Educational & Nutrition~\cite{buccinca2021trust}, Student Performance~\cite{rastogi2022deciding}, Analytical Reasoning (Text~\cite{he2023knowing, cau2023effects}), Admissions~\cite{ma2025towards} 
        \\
       
        Leisure & Bird Classification~\cite{morrison2024impact}, Maze Completion~\cite{vasconcelos2023explanations}, News Classification~\cite{pareek2024effect}, Trip Planning~\cite{salimzadeh2024dealing}, Logic Puzzles~\cite{swaroop2025personalising, swaroop2024accuracy}, Exercise Recommendation~\cite{buccinca2025contrastive}, Movie Reviews~\cite{lu2024does}
        \\ 
        
    \bottomrule

\end{tabular}
\end{table}

\begin{table}[htpb]
  \caption{A summary of intervention mechanics (datasets and ML models), decision-making types (binary and multi-class), and target population type. There is a diverse use of datasets. The majority of studies recruit crowd-workers.}
  \label{tab-data-model}
  \Description{Table 7. The table provides an overview of categories, data types, data classes, ML models, deep learning models, decision-making, and participants. The table has two columns labeled Category and Studies.}
  \begin{tabular}{@{}p{2.1cm} p{12.3cm}}
    \toprule
    Category & Studies  \\
    \midrule
       
        Participants & Users~\cite{cabitza2023ai}, Crowd-workers~\cite{buccinca2021trust, schemmer2023appropriate, schemmer2023towards, vasconcelos2023explanations, he2023knowing, he2023stated, cau2023effects, ma2024you, salimzadeh2024dealing, schoefer2024explanations, he2024err, morrison2024impact, swaroop2024accuracy, he2025conversational, swaroop2025personalising, buccinca2025contrastive}, General~\cite{yang2020how}
        
        \\
        
        Decisions & Binary~\cite{yang2020how, buccinca2021trust, schemmer2023appropriate, cabitza2023ai, he2023stated, cau2023effects, chiang2023two, ma2024you, schoefer2024explanations, he2024err, he2025conversational}, Multi-Class~\cite{schemmer2023towards, cabitza2023ai, vasconcelos2023explanations, he2023knowing, cau2023effects, salimzadeh2024dealing, morrison2024impact, swaroop2024accuracy, swaroop2025personalising, buccinca2025contrastive}, Value~\cite{cabitza2023ai}
        
        \\

        Dataset Type & COMPASS~\cite{chiang2023two}, Income~\cite{ma2024you}, Food Ingredients~\cite{buccinca2021trust}, Leaf~\cite{yang2020how}, Reviews and Text Classification~\cite{schemmer2023appropriate, he2023knowing, cau2023effects, schoefer2024explanations, he2024err, lu2024does}, Students~\cite{ma2025towards}, Loan~\cite{he2023stated, he2025conversational}, Medical (Images, X-Rays, ECG, MRI, Profiles, etc.)~\cite{cabitza2023ai, swaroop2024accuracy, swaroop2025personalising}, Natural or General Images~\cite{cau2023effects}, Birds~\cite{schemmer2023towards, morrison2024impact}, Job Applications~\cite{cau2025influence}
        \\
        
        ML Models & SVM~\cite{yang2020how, schemmer2023appropriate}, Logistic/Linear/OLS Regression~\cite{he2023stated, ma2024you, ma2025towards}, Gradient Boosting~\cite{he2025conversational, cau2025influence}, RoBERTa~\cite{lu2024does}, ResNet50~\cite{schemmer2023towards}, LogiFormer~\cite{he2023knowing}, CNN~\cite{cau2023effects}, BERT~\cite{he2024err}
        
        \\

    \bottomrule

\end{tabular}
\end{table}

\begin{table}[htbp]
  \caption{An overall summary of included studies. Studies mostly use empirical evaluations with the crowd-worker population, and favor the multi-stage decisions. Studies mostly explore binary or multi-class decisions. Diverse explanation methods are used with a high focus on example-based and feature-based XAI. Notes: Most reported attributes (\textbullet) are mutually non-exclusive.}
  \Description{Table 8. The table shows the overall division of studies. The table has columns Sr, Study Ref, Year, Study Type (subdivided into Empirical, Qualitative, Mixed), Target (subdivided into Users, Crow-workers, General), Decisions (subdivided into Single-stage, Multi-stage, Binary, Multi-class), Tasks (subdivided into Value, Other), and Explanations (subdivided into Model-based, Feature-based, Example-based, General).}
  \label{tab-study-main-table}
  
  \begin{tabular}{lllllllllllllllllll}
    \toprule
    ~ & ~ & ~ & \multicolumn{3}{c}{\textbf{Study Type}} & \multicolumn{3}{c}{\textbf{Target}} & \multicolumn{2}{c}{\textbf{Decisions}}  & \multicolumn{3}{c}{\textbf{Tasks}} & \multicolumn{4}{c}{\textbf{Explanations}} \\
    
    Sr & Study Ref & Year & \rotatebox[origin=b]{90}{Empirical} & \rotatebox[origin=b]{90}{Qualitative} & \rotatebox[origin=b]{90}{Mixed} & \rotatebox[origin=b]{90}{Users} & \rotatebox[origin=b]{90}{Crowd-workers} & \rotatebox[origin=b]{90}{General} & \rotatebox[origin=b]{90}{Single-stage} & \rotatebox[origin=b]{90}{Multi-stage} & \rotatebox[origin=b]{90}{Binary} & \rotatebox[origin=b]{90}{Multi-class} & \rotatebox[origin=b]{90}{Value} & \rotatebox[origin=b]{90}{Model-based} & \rotatebox[origin=b]{90}{Feature-based} & \rotatebox[origin=b]{90}{Example-based} & \rotatebox[origin=b]{90}{General} \\
    \midrule
    
        1 & Yang et al.~\cite{yang2020how} & 2020 & \textbullet & ~ & ~ & ~ & ~ & \textbullet & \textbullet & ~ & \textbullet & ~ & ~ & ~ & ~ & \textbullet & ~ \\ 
        2 & Bu\c{c}inca et al.~\cite{buccinca2021trust} & 2021 & \textbullet & ~ & ~ & ~ & \textbullet & ~ & ~ & \textbullet & \textbullet & ~ & ~ & ~ & \textbullet & ~ & ~ \\ 
        3 & Schemmer et al.~\cite{schemmer2023appropriate} & 2023 & \textbullet & ~ & ~ & ~ & \textbullet & ~ & ~ & \textbullet & \textbullet & ~ & ~ & ~ & \textbullet & \textbullet & ~ \\ 
        4 & Schemmer et al.~\cite{schemmer2023towards} & 2023 & \textbullet & ~ & ~ & ~ & \textbullet & ~ & ~ & \textbullet & ~ & \textbullet & ~ & ~ & ~ & \textbullet & ~ \\ 
        5 & Cabitza et al.~\cite{cabitza2023ai} & 2023 & \textbullet & ~ & ~ & \textbullet & ~ & ~ & ~ & \textbullet & \textbullet & \textbullet & \textbullet & ~ & \textbullet & ~ & ~ \\ 
        6 & Vasconcelos et al.~\cite{vasconcelos2023explanations} & 2023 & \textbullet & ~ & ~ & ~ & \textbullet & ~ & ~ & \textbullet & ~ & \textbullet & ~ & ~ & ~ & \textbullet & ~ \\ 
        7 & He et al.~\cite{he2023knowing} & 2023 & \textbullet & ~ & ~ & ~ & \textbullet & ~ & ~ & \textbullet & ~ & \textbullet & ~ & ~ & \textbullet & ~ & ~ \\ 
        8 & He et al.~\cite{he2023stated} & 2023 & \textbullet & ~ & ~ & ~ & \textbullet & ~ & ~ & \textbullet & \textbullet & ~ & ~ & \textbullet & ~ & ~ & ~ \\ 
        9 & Cau et al.~\cite{cau2023effects} & 2023 & \textbullet & ~ & ~ & ~ & \textbullet & ~ & ~ & \textbullet & \textbullet & \textbullet & ~ & ~ & \textbullet & \textbullet & ~ \\ 
        10 & Chiang et al.~\cite{chiang2023two} & 2023 & \textbullet & ~ & ~ & ~ & \textbullet & ~ & ~ & \textbullet & \textbullet & ~ & ~ & ~ & ~ & ~ & \textbullet \\ 
        11 & Ma et al.~\cite{ma2024you} & 2024 & \textbullet & ~ & ~ & ~ & \textbullet & ~ & ~ & \textbullet & \textbullet & ~ & ~ & \textbullet & \textbullet & ~ & ~ \\ 
        12 & Salimzadeh et al.~\cite{salimzadeh2024dealing} & 2024 & \textbullet & ~ & ~ & ~ & \textbullet & ~ & ~ & \textbullet & ~ & \textbullet & ~ & ~ & ~ & ~ & \textbullet \\ 
        13 & Schoeffer et al.~\cite{schoefer2024explanations} & 2024 & \textbullet & ~ & ~ & ~ & \textbullet & ~ & \textbullet & ~ & \textbullet & ~ & ~ & ~ & \textbullet & ~ & ~ \\ 
        14 & He et al.~\cite{he2024err} & 2024 & \textbullet & ~ & ~ & ~ & \textbullet & ~ & ~ & \textbullet & \textbullet & ~ & ~ & ~ & \textbullet & ~ & ~ \\ 
        15 & Morrison et al.~\cite{morrison2024impact} & 2024 & ~ & ~ & \textbullet & ~ & \textbullet & ~ & ~ & \textbullet & ~ & \textbullet & ~ & ~ & ~ & \textbullet & \textbullet \\ 
        16 & Swaroop et al.~\cite{swaroop2024accuracy} & 2024 & \textbullet & ~ & ~ & ~ & \textbullet & ~ & \textbullet & \textbullet & ~ & \textbullet & ~ & ~ & ~ & ~ & \textbullet \\ 
        17 & Lu et al.~\cite{lu2024does} & 2024 & \textbullet & ~ & ~ & ~ & \textbullet & ~ & ~ & \textbullet & \textbullet & ~ & ~ & ~ & ~ & ~ & \textbullet \\ 
        18 & He et al.~\cite{he2025conversational} & 2025 & \textbullet & ~ & ~ & ~ & \textbullet & ~ & ~ & \textbullet & \textbullet & ~ & ~ & ~ & \textbullet & ~ & ~ \\ 
        19 & Swaroop et al.~\cite{swaroop2025personalising} & 2025 & \textbullet & ~ & ~ & ~ & \textbullet & ~ & \textbullet & \textbullet & ~ & \textbullet & ~ & ~ & ~ & ~ & \textbullet \\ 
        20 & Bu\c{c}inca et al.~\cite{buccinca2025contrastive} & 2025 & \textbullet & ~ & ~ & ~ & \textbullet & ~ & ~ & \textbullet & ~ & \textbullet & ~ & ~ & ~ & ~ & \textbullet \\ 
        21 & Cau and Spano~\cite{cau2025influence} & 2025 & \textbullet & ~ & ~ & ~ & \textbullet & ~ & ~ & \textbullet & \textbullet & ~ & ~ & \textbullet & \textbullet & ~ & ~ \\ 
        22 & Ma et al.~\cite{ma2025towards} & 2025 & ~ & ~ & \textbullet & ~ & \textbullet & ~ & ~ & \textbullet & \textbullet & ~ & ~ & ~ & \textbullet & ~ & ~ \\ 
        
        \midrule
        ~ & Total & 22 & 20 & 0 & 2  & 1 & 20 & 1 & 4 & 20 & 14 & 10 & 1 & 3 & 11 & 6 & 6 \\ 
        
        \bottomrule
  \end{tabular}
\end{table}

\end{document}